\documentclass[a4paper,12pt]{article}
\usepackage{graphicx,rotating,hyperref,slashed,amsmath,charter,xcolor,catchfilebetweentags,ifluatex,cite}
\makeatletter
\hypersetup{colorlinks,bookmarksopen,bookmarksnumbered,
linkcolor=blus,pdfstartview=FitH,urlcolor=rossos,citecolor=verde}

 \renewcommand{\Im}{{\rm Im}\,}
\renewcommand{\Re}{{\rm Re}\,}
\newcommand{\mio}[1]{}
\newcommand{\xxx}[1]{{\color{red}[\bf #1]}}

\newcommand{\etat}{\eta}
\newcommand{\bra}[1]{\langle #1 |}
\newcommand{\ket}[1]{| #1 \rangle}
\newcommand{\bk}[2]{\langle #1  |  #2  \rangle}
\newcommand{\kb}[2]{| #1  \rangle\langle   #2  |}
\newcommand{\bAk}[3]{\langle #1  |  #2|#3  \rangle}
\definecolor{Gray}{gray}{0.95}
\newcommand{\bbox}[1]{\fcolorbox{gray}{Gray}{~$\displaystyle #1$~}}
\newcommand{\psit}{\psi}

\newcommand{\sfrac}[2]{#1/#2}

\usepackage{multicol}
\usepackage{color}
\definecolor{rosso}{cmyk}{0,1,1,0.4}
\definecolor{rossos}{cmyk}{0,1,1,0.55}
\definecolor{rossoc}{cmyk}{0,1,1,0.2}
\definecolor{blu}{cmyk}{1,1,0,0.3}
\definecolor{blus}{cmyk}{1,1,0,0.6}
\definecolor{bluc}{cmyk}{1,1,0,0.1}
\definecolor{verde}{cmyk}{0.92,0,0.59,0.25}
\definecolor{verdec}{cmyk}{0.92,0,0.59,0.15}
\definecolor{verdes}{cmyk}{0.92,0,0.59,0.4}

\newcommand{\eq}[1]{~{\rm (\ref{eq:#1})}}

\def\circa#1{\,\raise.3ex\hbox{$#1$\kern-.75em\lower1ex\hbox{$\sim$}}\,}

\newcommand{\beq}{\begin{equation}}
\newcommand{\eeq}{\end{equation}}
\newcommand{\mb}[1]{\mbox{\boldmath $#1$}}

\newcommand{\bea}{\begin{eqnarray}}
\newcommand{\eea}{\end{eqnarray}}
\newcommand{\be}{\begin{equation}}
\newcommand{\ee}{\end{equation}}
\font\tenrsfs=rsfs10 at 12pt
\font\sevenrsfs=rsfs7 at 10 pt
\font\fiversfs=rsfs5
\newfam\rsfsfam
\textfont\rsfsfam=\tenrsfs
\scriptfont\rsfsfam=\sevenrsfs
\scriptscriptfont\rsfsfam=\fiversfs
\def\mathscr#1{{\fam\rsfsfam\relax#1}}

\def\Lag{\mathscr{L}}

\def\circa#1{\,\raise.3ex\hbox{$#1$\kern-.75em\lower1ex\hbox{$\sim$}}\,}
\makeatletter

\def\hhref#1{\href{http://arxiv.org/abs/#1}{arXiv:#1}} 

\def\hhref#1{\href{http://arxiv.org/abs/#1}{arXiv:#1}} 
 
\def\art{\@ifnextchar[{\eart}{\oart}}
\def\eart[#1]#2#3#4#5#6{{\rm #2}, {\em #3 \bf #4} {\rm (#6) #5} ({\em #1})}

\def\article{\@ifnextchar[{\earticle}{\oarticle}}
\def\oarticle#1#2#3#4#5#6{{\rm #1}, {\em ``#6''}, {\rm #2 #3 (#5) #4}}
\def\earticle[#1]#2#3#4#5#6#7{{\rm #2}, {\em ``#7''}, {\rm #3 #4 (#6) #5}  [\hhref{#1}]}
\def\hepart[#1]#2{{\rm #2, \em#1}}
\def\heparticle[#1]#2#3{#2, {\em ``#3''} [\hhref{#1}]}

\newcounter{alphaequation}[equation]
\def\thealphaequation{\theequation\hbox to
0.6em{\hfil\alph{alphaequation}\hfil}}
\def\eqnsystem#1{
\def\@eqnnum{{\rm (\thealphaequation)}}
\def\@@eqncr{\let\@tempa\relax \ifcase\@eqcnt \def\@tempa{& & &} \or
  \def\@tempa{& &}\or \def\@tempa{&}\fi\@tempa
  \if@eqnsw\@eqnnum\refstepcounter{alphaequation}\fi
\global\@eqnswtrue\global\@eqcnt=0\cr}
\refstepcounter{equation} \let\@currentlabel\theequation \def\@tempb{#1}
\ifx\@tempb\empty\else\label{#1}\fi
\refstepcounter{alphaequation}
\let\@currentlabel\thealphaequation
\global\@eqnswtrue\global\@eqcnt=0 \tabskip\@centering\let\\=\@eqncr
$$\halign to \displaywidth\bgroup \@eqnsel\hskip\@centering
$\displaystyle\tabskip\z@{##}$&\global\@eqcnt\@ne
\hskip2\arraycolsep\hfil${##}$\hfil& \global\@eqcnt\tw@\hskip2\arraycolsep
$\displaystyle\tabskip\z@{##}$\hfil
\tabskip\@centering&\llap{##}\tabskip\z@\cr}
\def\endeqnsystem{\@@eqncr\egroup$$\global\@ignoretrue} \makeatother

\begin{document}

\vspace{1cm}

\begin{center}
{\LARGE \bf \color{rossos}
Quantum mechanics of 4-derivative theories}\\[1cm]

{\large\bf Alberto Salvio$^{a}$ {\rm and} Alessandro Strumia$^{b}$}  
\\[7mm]
{\it $^a$ } {\em Departamento de F\'isica Te\'orica, Universidad Aut\'onoma de Madrid\\ and Instituto de F\'isica Te\'orica IFT-UAM/CSIC,  Madrid, Spain}\\[3mm]
{\it $^b$ Dipartimento di Fisica dell'Universit{\`a} di Pisa and INFN, Italia\\
CERN, Theory Division, Geneva, Switzerland}

\vspace{1cm}
{\large\bf\color{blus} Abstract}
\begin{quote}
A renormalizable theory of gravity is obtained if 
the dimension-less 4-derivative kinetic term of the graviton, 
which classically suffers from negative unbounded energy,
admits a sensible quantization.
We find that a 4-derivative degree of freedom involves a canonical coordinate
with unusual time-inversion parity, 
and that a correspondingly unusual representation 
must be employed  for the relative quantum operator.
The resulting theory has positive energy eigenvalues, normalizable wave functions,
unitary evolution in a negative-norm configuration space.
We present a formalism for quantum mechanics with a generic norm.

\end{quote}

\thispagestyle{empty}
\end{center}
\begin{quote}
{\large\noindent\color{blus} 
}

\end{quote}
\tableofcontents

\setcounter{footnote}{0}

\newpage


\section{Introduction}

Newton invented classical mechanics putting two time derivatives in his equation  $F=m \ddot x$,
which corresponds to a kinetic energy with 2 time derivatives, $m\dot x^2/2$.
Later Ostrogradski proofed a no-go theorem: non-degenerate classical systems with more than two time derivatives contain arbitrarily negative
energies and develop fatal run-away instabilities~\cite{Ostro}. 
Classically, they do not make sense.

The discovery that nature is relativistic and quantum  opened the quest for an extension of Newtonian gravity.
A century ago Einstein and Hilbert found the classical theory of relativistic gravity.
However, its quantum version is not renormalizable in 3+1 space-time dimensions.
Sticking to the observed number of space-time dimensions, a
renormalizable extension of general relativity is found
by adding terms quadratic in the curvature tensor to the Einstein-Hilbert Lagrangian, such that
the graviton acquires a 4-derivative kinetic term.
Stelle proposed and dismissed this extension 
\mio{commenting that ``difficulties with unitarity appear to preclude their direct acceptability as physical theories''} \cite{Stelle} (see also~\cite{Frad,Avramidi,Antoniadis,EOR,Shapiro,agravity,Einhorn}).

Recently the Higgs mass hierarchy problem brought interest to dimension-less theories.
In this context, gravitons must have dimension 0 (being a dimension-less metric) and thereby {\em must} have a 4-derivative kinetic term.
If these theories could make sense at quantum level, despite the negative classical energy, 
a great deal would be gained: relativistic quantum gravity, plus hierarchies among dynamically generated mass scales~\cite{agravity}, plus inflation~\cite{ainflation,Farzinnia:2015fka}.

\medskip


Quantization can eliminate arbitrarily negative classical energies.
The following example  is well known:
the classical relativistic spin 1/2 field is described by a spinor $\Psi(x)$ with 
Dirac  Lagrangian $\Lag = \bar \Psi (i\slashed{\partial} - m)\Psi$ containing one time derivative.
Treating $\Psi$ as a classical field (as initially proposed by Schr\"{o}dinger),
and inserting the plane-wave expansion
\beq\Psi(x) = \int \frac{d^3 p}{(2\pi)^3 \sqrt{2E_p}} [a_{p,s} u_{p,s} e^{-i p\cdot x}  + b^\dagger_{p,s} v_{p,s} e^{ip\cdot x} ]\eeq
in the Hamiltonian 
one finds negative energies in half of the configurations space:\footnote{Unlike in the case of the Hydrogen atom emphasized by Woodard~\cite{Woodard}, 
where the instability eliminated by quantum mechanics
occurs only in one point of the configuration space.}
\beq H \mio{\,= \int d^3 x \,\bar\Psi [m-i \mb{\gamma}\cdot\mb{\nabla}] \Psi }
= \int \frac{d^3 p}{(2\pi )^3} \,  E_p   [a_{p,s}^\dagger a_{p,s} - b_{p,s}b_{p,s}^\dagger ],\qquad
E_p = \sqrt{m^2+\vec p^2}
\eeq
This classical arbitrarily negative energy is avoided by
quantization with anti-commutators\mio{\footnote{Observables, built with pairs of fermions fields, commute at space-like distances.}}, if the vacuum state is appropriately chosen.
Indeed, the two-state solution to $\{ b,b^\dagger\}=1$ shows that one can switch annihilation with creation operators by choosing 
the vacuum  to be the state with lower energy.
\mio{With two states, this amounts to
\beq b = \bordermatrix{ & \ket{0} & \ket{1} \cr
\bra{0} & 0&1\cr  
\bra{1} &0 &0 } = \bordermatrix{ & \ket{1} & \ket{0} \cr
\bra{1} & 0&0\cr  
\bra{0} &1 &0 } = \tilde b^\dagger.
\eeq}

The spin 0 and spin 1 relativistic fields (described by dimension-1 fields with 2 derivatives) do not have this issue: 
the negative-frequency solutions to the Klein-Gordon equation correspond to   Hamiltonians with positive energy.\mio{\footnote{Various text-books incorrectly draw analogies  with the Dirac field.}}

The general lesson is that quantization depends on the number of time derivatives.

\medskip

The goal of this study is describing if/how systems with 4 derivatives can be quantized obtaining a consistent theory,
in particular of quantum gravity.  We will find that a unique structure emerges, that again involves switching annihilation and creation operators.

\medskip

This will bring us into the territory of negative norm quanta,
avoided like a plague by serious theorists that call them `ghosts',
and explored only by notorious crackpots such as Dirac~\cite{Dirac}, Pauli~\cite{Pauli}, Heisenberg~\cite{Heisenberg},\mio{\footnote{The original physical motivation  of these studies is dead.}}
Pais, Uhlenbeck~\cite{Pais}, Lee, Wick~\cite{LW}, Cutkosky~\cite{Cutkosky}, Coleman~\cite{Coleman}, Feynman~\cite{Feynman}, Gross~\cite{Gross},
Hawking~\cite{HH}, 't Hooft~\cite{tHooft} and others, also more recently~\cite{Grinstein,BM,Smilga,Gonera,Biswas:2011ar,Chen,BRST,Maldacena,Lu:2011ks}. 
These works sometimes contain  bizarre and confusing statements
and obsolete motivations, together with  interesting ideas and ad-hoc prescriptions.

\medskip

This paper is structured as follows.
In section~\ref{OCO} we review the canonical Ostrogradski formalism.
In section~\ref{QM} we present negative norm quantum mechanics,
 the negative norm harmonic oscillator (section~\ref{osc-}), 
 and the associated  negative-norm representation of a canonical coordinate (section~\ref{secDP}),
with unusual parity under time-inversion $T$. 
 In section~\ref{DP} we recall that
a 4-derivative degree of freedom $q(t)$ is described by two canonical coordinates: $q_1 = q$ and $q_2 = \dot q$.
While $q_1$ is $T$-even as usual, $q_2$ is $T$-odd: we argue that thereby it naturally follows the negative-norm representation.
The resulting quantum theory is unitary: time evolution preserves the negative norm.
The path-integral formulation is discussed in section~\ref{pathI}.
In section~\ref{int} we discuss the interacting theory, outline the extension to quantum field theory, and discuss the issue of
giving a sensible interpretation to negative norms, via a postulate that generalizes the Born rule.
Conclusions are given in section~\ref{end}.

\section{The Ostrogradski classical canonical  formalism}\label{OCO}
Let us now introduce the main issues in the simplest relevant case.
Our final goal will be 4-derivative gravity; however
the graviton components
can be Fourier expanded into modes with given momentum and 4 time derivatives,
and at leading order in the perturbative expansion one has decoupled harmonic oscillators.
So, we start considering a single mode $q(t)$, described by the Lagrangian
\beq \label{eq:LagO}
\Lag =  -\frac{\ddot q^2}{2} + (\omega_1^2+\omega_2^2) \frac{\dot q^2}{2} - \omega_1^2 \omega_2^2 \frac{q^2}{2}-V(q)
=-\frac12  q (\frac{d^2}{dt^2} + \omega_1^2)(\frac{d^2}{dt^2} + \omega_2^2)q - V(q) + \mbox{ total derivatives}.
\nonumber \eeq
where $V(q)$ is some interaction. 
We assume real $\omega_{1}$, $\omega_{2}$, because we are interested in ghosts (negative kinetic and potential energy), not in tachyonic instabilities
(potential unstable with respect to the kinetic term).
The $-$ sign means that the ghost is the state with larger $\omega$;
we choose $\omega_1> \omega_2$ and don't explicitly discuss  here the degenerate case $\omega_1=\omega_2$.
 
 \smallskip
 
 Ostrogradski introduced an {\em auxiliary} coordinate $q_2$ that allows to describe the 4-derivative oscillator in canonical Hamiltonian form (see also ref.~\cite{Pais} for a review of this method):
\beq \label{eq:Ostroq2}
\begin{array}{ll}
q_1 = q,\qquad  &\displaystyle 
p_1 = \frac{\delta   \Lag}{\delta \dot q_1}  =(\omega_1^2+\omega_2^2)\dot q + \dddot q ,\\[5mm]
q_2 =\lambda  \dot q,&\displaystyle p_2 = \frac{\delta   \Lag}{\delta \dot q_2} = -\frac{\ddot q}{\lambda},
\end{array}\eeq
 where 
for a generic variable $x$ we have introduced the variational derivative
\be \frac{\delta \Lag}{\delta x} = \frac{\partial\Lag}{\partial x} - \frac{d}{dt} \frac{\partial \Lag}{\partial \dot x} +\frac{d^2}{dt^2} \frac{\partial \Lag}{\partial \ddot x} + \cdots.  \ee While Ostrogradski assumed $\lambda=1$, we introduced an arbitrary constant $\lambda$.\mio{\footnote{One can alternatively use
$\ddot q$ rather than $\dot q$ as auxiliary coordinate: this just amount to switch $q_2 \leftrightarrow p_2$ with respect to the Ostrogradski choice.}}
  The system in eq.~(\ref{eq:Ostroq2}) can be solved for $q$ and its time derivatives,
\beq \label{eq:Ostroq3}
q = q_1,\qquad    
\dot q = \frac{q_2}{\lambda}, \qquad
\ddot q = - \lambda  p_2,\qquad \dddot q = p_1-\left(\omega_1^2+\omega_2^2\right) \frac{q_2}{\lambda},
\eeq
and the Hamiltonian turns out to be
\beq \label{eq:HOstro}
H = \sum_{i=1}^2 p_i \dot q_i - \mathscr{L}=
\frac{p_1 q_2}{\lambda} - \frac{\lambda^2}{2}p_2^2 - \frac{\omega_1^2+\omega_2^2}{2\lambda^2} q_2^2 +\frac{\omega_1^2\omega_2^2}{2}q_1^2+V(q_1).\eeq
In view of its first term, the classical Hamiltonian $H$ has no minimal energy configuration: this is the essence of the Ostrogradski no-go classical theorem.
Using the Poisson parentheses $\{~,~\}$ one computes the Hamiltonian equations of motion:
\beq \left\{\begin{array}{ll}\displaystyle
\dot q_1 = \{q_1,H\} = \frac{\partial H}{\partial p_1} =\frac{q_2}{\lambda},\qquad  &\displaystyle 
\dot p_1 =  \{p_1,H\} = -\frac{\partial H}{\partial q_1}=  -\omega_1^2\omega_2^2 q_1 - V'(q_1) ,\\[5mm]  \displaystyle
\dot q_2 =  \{q_2,H\} = \frac{\partial H}{\partial p_2}=-\lambda^2 p_2,&\displaystyle 
\dot p_2 = \{p_2,H\} = -\frac{\partial H}{\partial q_2}=-\frac{p_1}{\lambda} + \frac{\omega_1^2+\omega_2^2}{\lambda^2} q_2.
\end{array}\right.\eeq
For any $\lambda$ they imply the classical Lagrangian equation of motion.
Setting $V=0$, it is
\beq \label{eq:classq''''}
(\frac{d^2}{dt^2} + \omega_1^2)(\frac{d^2}{dt^2} + \omega_2^2)q=
\frac{d^4 q}{dt^4}+ (\omega_1^2 +\omega^2_2)\frac{d^2q}{dt^2} + \omega_1^2 \omega_2^2 q=0.\eeq
The corresponding classical solution,
for given initial conditions $q_0, \dot q_0, \ddot q_0, \dddot{q}_0$ at $t=0$,
is
\beq \label{eq:q(t)}
q(t) = -\frac{\omega_2^2 q_0 + \ddot q_0}{\omega_1^2-\omega_2^2} \cos(\omega_1 t) + 
\frac{\omega_1^2 q_0 + \ddot q_0}{\omega_1^2-\omega_2^2} \cos(\omega_2 t) -
\frac{\omega_2^2 \dot q_0 + \dddot q_0}{\omega_1(\omega_1^2-\omega_2^2)} \sin(\omega_1 t) +
\frac{\omega_1^2 \dot q_0 + \dddot q_0}{\omega_2(\omega_1^2-\omega_2^2)} \sin(\omega_2 t) .
\eeq
This is a well behaved oscillator without run-away issues
because the positive-energy and negative-energy components are decoupled.
Run-away solutions appear when they interact through a generic interaction, such as a $V\neq 0$.

\subsection{Quantizing the Ostrogradski Hamiltonian}\label{OCO2}
The classical equation differs from the usual 2-derivative equation $ d(q+ip)/dt = i\omega (q+ip)$, so that, trying to quantize the theory, 
we do not define the usual annihilation operator $a_i \propto q_i + i p_i $.
Rather, it is convenient to define the operators $a_i$ as the coefficients of  given frequency:
\beq q(t) = a_1 e^{-i\omega_1 t} + a_2 e^{-i\omega_2 t} + \hbox{h.c.}\eeq
The $a_1,a_2$ can be expressed in terms of canonical Hamiltonian coordinates:
\bea
a_1 &=& \frac{\lambda \omega_1 \omega_2^2 q_1 - i \omega_1^2 q_2 + i p_1  \lambda- \omega_1 p_2 \lambda^2}{2\lambda \omega_1(\omega_2^2-\omega_1^2)}, \label{a1}\\
a_2 &=&\frac{\lambda\omega_1^2 \omega_2 q_1 - i \omega_2^2 q_2 + i p_1 \lambda - \omega_2p_2 \lambda^2}{2\lambda\omega_2(\omega_1^2-\omega_2^2)}. \label{a2}
\eea
Using the canonical quantization $[q_i,p_j]=i\delta_{ij}$ one finds the commutation relations for the $a_i$:
\beq \label{eq:a12comm}
[\tilde a_1, \tilde  a^\dagger_1]=-1,\qquad
 [\tilde a_2, \tilde  a^\dagger_2]=1,\qquad \hbox{all other commutators vanish.}\eeq
having normalised $\tilde a_1 = \sqrt{2\omega_1(\omega_1^2-\omega_2^2)}a_1$ and 
$\tilde a_2 = \sqrt{2\omega_2(\omega_1^2-\omega_2^2)}a_2$.
The Hamiltonian is 
\beq \label{eq:Htilde}
H=- \omega_1  \frac{ \tilde a_1 \tilde a_1^\dagger+ \tilde a_1^\dagger \tilde a_1}{2}   +    \omega_2 \frac{\tilde a_2 \tilde a_2^\dagger+\tilde a_2^\dagger \tilde a_2}{2}. \eeq
The state 1 with higher frequency $\omega_1>\omega_2$ is a ghost.

\medskip

As better discussed later in section~\ref{osc-}, this system can be quantized in two different ways:
\begin{itemize}
\item[1.] {\bf Positive norm, negative energy}.
One redefines $\tilde a'_1=\tilde a_1^\dagger$, such that it has the usual commutation  $[\tilde a'_1,\tilde a^{\prime\dagger}_1]=1$.
The vacuum state $\ket{\tilde 0}$ is defined as usual by $\tilde a'_1\ket{\tilde 0}=0$ and $\tilde a_2\ket{\tilde 0}=0$.
By solving this condition as a differential equation for $\psi_{\tilde 0}(q_1, q_2) = \bk{q_1,q_2}{\tilde 0}
$ with $p_i = - i \partial/\partial q_i$
one obtains the ground state wave function:
\beq \psi_{ \tilde 0}(q_1,q_2) = \exp\bigg(-\frac{q_1^2 \omega_1 \omega_2  +  q_2^2/\lambda^2}{2} (\omega_1 - \omega_2) + i  q_1 \frac{q_2}{\lambda}\omega_1 \omega_2\bigg).\eeq

\item[2.] {\bf Negative norm, positive energy}.
The vacuum is now defined as  $a_1\ket{ 0}=0$ and $a_2\ket{ 0}=0$.
Using $p_i = - i \partial/\partial q_i$ one obtains the ground-state wave function 
\beq\label{eq:psi0div}
 \psi_{ 0}(q_1,q_2) \propto \exp\bigg(\frac{-q_1^2 \omega_1 \omega_2 +   q_2^2/\lambda^2}{2}  (\omega_1 + \omega_2)  - i  q_1 \frac{q_2}{\lambda}\omega_1 \omega_2\bigg).\eeq

\end{itemize}
If $\lambda=1$ the situation is bad, as emphasized by~\cite{Woodard}: the positive-norm quantization gives
a normalizable wave-function $\psi_{\tilde 0}$ but negative energies;
the negative-norm quantization gives a ground state wave function not normalizable in $q_2 = \dot q$.
Excited states have the same problem.

\smallskip

However, as we will show in section~\ref{DP}, consistency demands 
the negative-norm Dirac-Pauli representation of a canonical coordinate 
which roughly amounts to choosing an imaginary $\lambda$, e.g.\ $\lambda=- i$.
One then obtains positive energy, negative norm, and a wave function $\psi_0(q_1,q_2)$ normalizable in $q_1$ and 
$q_2 =- i \dot q$.
As we will now discuss, despite the strange $i$ factor, $\dot q=iq_2$ as well as  the
Ostrogradski Hamiltonian $H = i  q_2 p_1 +\cdots = \dot q p_1+\cdots $ are self-adjoint,
so that time evolution is unitary.

\medskip



\section{Quantum mechanics with negative norm}\label{QM}
We here discuss quantum mechanics with negative norm from a general point of view.
Negative norm states require putting some minus sign here and there.
It is convenient to be more general and consider 
a Hilbert-like space with generic, possibly negative, constant norm 
(called Krein space by mathematicians)
and develop a basis-independent formalism.
This will let us to clarify confusions, in particular about self-adjoint operators
that are represented (in some basis) by non-hermitian matrices, allowing us to understand the unusual imaginary $\lambda$ introduced in the previous section.

\medskip

We follow the notations used in general relativity, rewriting the quantum state
metric as  $\bk{_n}{_m}=\eta_{nm}$ and defining the inverse metric  
 $  (\etat)^{nm}\equiv(\eta^{-1})_{nm}$,
 the contro-variant ket  $\ket{^n}=\etat^{nm}\ket{_m}$ such that
$ \bk{^n}{^m}=\etat^{nm}$ and $\bk{^n}{_m}=\delta^n_m = \bk{_n}{^m}$.
Summations over repeated indexes are implicit.
As usual, bras denote complex conjugate of kets.

 A generic {\em state} $\ket{\psi}$ can be expanded in either the `covariant'  or the
`controvariant' basis:
 \beq 
\label{eq:psicocov} \psi_n \equiv \bk{_n}{\psi},\qquad \psit^n \equiv \bk{^n}{\psi}.\eeq
Then 
\beq
\ket{\psi} =  \psit^n \ket{_n} = \psi_n \ket{^n}.\eeq
A generic {\em linear operator} $A$ can be written as a matrix in 4 different ways: 
\beq
A_{nm}\equiv \bAk{_n}{A}{_m},\qquad
A^{nm}\equiv\bAk{^n}{A}{^m},\qquad
A_n{}^m\equiv \bAk{_n}{A}{^m},\qquad
A^n{}_m\equiv \bAk{^n}{A}{_m}.\eeq
Then
\beq A ={A}^{nm}\ket{_n}\bra{_m} 
={A}_{nm}\ket{^n}\bra{^m}
={A}_{n}{}^{m}\ket{^n}\bra{_m}
={A}^n{}_m\ket{_n}\bra{^m}
.\eeq
The components of the matrices are related by $A_n{}^m = \eta_{nn'} A^{n'}{}_{m'}\etat^{m'm}$
which is an {\em iso-spectral} transformation:
the eigenvalues do not change because the matrix $A$ gets left-multiplied by $\eta$ and right-multiplied by its inverse.

The {\em unity operator} is represented by $1_{nm}=\eta_{nm}$ and ${1}^{nm}={\etat}^{nm}$
and expanded as
 \beq 1 = \etat^{nm} |_n\rangle\langle _m | =\eta_{nm}\kb{^n}{^m} = \kb{_n}{^n}=\kb{^n}{_n}   .\eeq
Operator multiplication becomes, in components, $(AB)_{nm} = A_{nn'} \etat^{n'm'} B_{m'm}$.
Expectation values are given by $\bAk{\psi}{A}{\psi}/\bk{\psi}{\psi}$.

\smallskip

The {\em adjoint} $A^\dagger$ of an operator $A$ is defined, as usual, as the operator such that
$\ket{\psi'} = A \ket{\psi}$ implies $\bra{\psi'} = \bra{\psi} A^\dagger$.
Thereby for generic matrix elements one has
$\bAk{\psi_2}{A^\dagger}{\psi_1} \equiv \bAk{\psi_1}{A}{\psi_2}^*$,
and  the relation for the components
\beq (A^\dagger)_{nm} = A_{mn}^*\qquad\hbox{i.e.}\qquad
 ( A^\dagger)^{nm} =  A^{mn*}\qquad\hbox{i.e.}\qquad
 ( A^\dagger)_n{}^m =  (A^m{}_n)^*. \label{eq:adj}
\eeq
The covariant components of a self-adjoint operator $A$ satisfy the usual condition:
a self-adjoint operator is described by a hermitian matrix, $A_{nm}$.
The same result holds for the contro-variant matrix $A^{nm}$.
The mixed components satisfy a different condition, where 
complex conjugation is supplemented by an iso-spectral transformation:
$A_n{}^m = (\eta A^{*T} \eta^{-1})_{n}{}^m$.\footnote{Here  $T$ denotes the matrix transpose; $*$ denotes complex conjugation;
$\dagger$ denotes the adjoint operation, that generalizes the usual 
hermitian conjugation and reduces to it in the positive norm case.
We never use $\dagger$ to denote hermitian conjugation of a matrix.}

A {\em self-adjoint} operator, $A^\dagger=A$  has real expectation values $\bAk{\psi}{A}{\psi}/\bk{\psi}{\psi}$,
  although the matrix $A^m{}_n$ that represents it can be anti-hermitian.


\smallskip

The mixed components directly enter into the {\em eigenvector equation} $A|\psi\rangle = A_\psi |\psi\rangle$:
\beq A_n{}^{m} \psi_m=
A_{nm'}\etat^{m'm} \psi_{m} = A_\psi \psi_n   \qquad \hbox{or}\qquad  
A^n{}_m\psit^m=
 \etat^{nn'} A_{n'm}\psit^m = A_\psi\psit^n\eeq
 where $A_\psi$ is the eigenvalue.
Let us now consider a self-adjoint operator $H$ (later it will be the Hamiltonian),
with eigenstates $\ket{E_n}$ and eigenvalues $E_n$.
The identity 
\beq \langle E_n | H |E_m  \rangle = \langle E_n | E_m\rangle E_m = E^*_n \langle E_n | E_m\rangle \eeq
tells that $H$  can have three different kinds of eigenstates:
\begin{itemize}
\item[$+$)] orthogonal eigenstates $\bk{E_n}{E_m}=0$
with real $E_n$ and norm $\bk{E_n}{E_n}= +1$;

\item[$-$)] orthogonal eigenstates $\bk{E_n}{E_m}=0$
with real $E_n$ and norm $\bk{E_n}{E_n}=- 1$;

\item[0)] pairs of complex conjugated eigenvalues, $E_n = E_m^*$
with $\bk{E_n}{E_m}\neq 0$ and zero norm, $\bk{E_n}{E_n}=0$.
\end{itemize}
In the classical analogue, the latter possibility corresponds to a ghost which is also a tachyon, 
which is a different kind of instability, to be avoided even in absence of ghosts.

\medskip

It is often convenient to choose a basis of eigenstates of $H$: $\ket{_n}=\ket{E_n}$.
The associated contro-variant states $\ket{^n}$ then satisfy $H\ket{^n} = E_n^*\ket{^n}$. 
In this basis the  space splits into two sectors: positive norm and negative norm,
plus the possible pairs of zero-norm states.
The two sectors experience a joint dynamics only if the initial state has a quantum entanglement among them.



\subsection{Unitary time evolution}\label{Unitarity}
The {\em evolution equation} $i \partial_t \ket{\psi} = H \ket{\psi}$ becomes
\beq  i \frac{\partial}{\partial t} \psi_n =  H_{nm}\etat^{mm'}  \psi_{m'}\qquad \hbox{or}\qquad
i \frac{\partial}{\partial t} \psit^n = \etat^{nn'} H_{n'm}  \psit^m .\eeq
The norm of any state $\ket{\psi(t)}$ is conserved by time evolution if $H$ is self-adjoint:
\beq i \frac{\partial}{\partial t} \bk{\psi'(t)}{\psi(t)} = \bAk{\psi'}{H-H^\dagger }{\psi}=0.\eeq
A self-adjoint Hamiltonian $H$ leads to unitary time evolution.
The explicit solution can be written as  $\ket{\psi(t)} = U(t) \ket{\psi(0)}$ with $U(t)={\rm T} e^{-i\int H(t) \,dt}$,
where T is the usual time-ordering.
In components,
\beq \psi_n(t) = U_n{}^m\psi_m(0)=
U_{nm'}\etat^{m'm} \psi_{m'}(0)      \qquad \hbox{or}\qquad       
\psit^n (t) =U^n{}_m\psi^m(0)= \etat^{nn'} U_{n'm}  \psit^m(0).\eeq

\bigskip

Having written generic-metric quantum mechanics in an abstract formalism that resembles as much as possible the usual
positive-norm formalism, let us now emphasize the key differences.
For simplicity, let us consider
 a time-independent $H$. One can then expand $U =  e^{-iHt} = \sum_{n=0}^\infty (-iHt)^n/n!$.
 
 \begin{itemize}
 \item Writing $U$ in mixed components, $U_n{}^m$ is the naive exponentiation of the matrix $H_n{}^m$.
 However, the mixed components of a self-adjoint $H$ do not form a  hermitian matrix.
Rather, the self-adjoint condition in eq.\eq{adj} dictates that they are hermitian up to an iso-spectral transformation.

\item The covariant components of a self-adjoint $H$ satisfy the usual Hermiticity $H^*_{nm}=H_{mn}$.
However, the covariant components
$U_{nm}$ are not given by the naive matrix exponentiation of $H_{nm}$.\mio{\footnote{Namely, by the {\tt MatrixExp} operation in the computer code {\sc Mathematica}.}}
 Rather, extra metric factors appear to covariantize the expansion:
 \beq  \label{eq:Ucov} 
U_{nm} = \eta_{nm} +  \eta_{nn'}  (-i Ht)^{n'm'} \eta_{m'm} + \frac12 \eta_{nn'}  (-i Ht)^{n'r'} \eta_{r's'} (-i Ht)^{s'm'} \eta_{m'm} + \cdots 
\eeq
Correspondingly, the unitarity condition $U^\dagger U  =1$ written in
covariant components is $ U_{n'n}^*\etat^{n'm'}  U_{m'm} = \eta_{nm}$,
while in mixed components one gets the usual
$U^{k*}_{~\,n} \, U_k{}^m=\delta_m^n$.
\end{itemize}
\mio{Multiplying eq.\eq{Ucov} at the left or at the right with the inverse metric, one recovers the usual matrix exponentiation for the mixed components
\beq U^n{}_m= \etat^{nn'} U_{n'm}  = e^{-i H^n{}_m t}\qquad\hbox{ and }\qquad 
U_n{}^m=U_{nm'}\etat_{m'm} = e^{-i H_n{}^m t}.\eeq}


\medskip

Practical computations often employ perturbation theory, which can now be 
easily generalized to generic norm.
Decomposing $H=H_0+V(t)$, 
the $I$nteraction picture is related to the Schroedinger picture as
$A_I = e^{iH_0 t} A e^{-iH_0 t}$ where
$A$ is any operator (including $V$).
Time evolution is given by
\beq  \label{eq:UI}
U_I (t_i,t_f)= {\rm T}\, e^{-i\int_{t_i}^{t_f} dt \,V_I(t)} = 1-i\int_{t_i}^{t_f} dt'\,V_I(t')-
\int_{t_i}^{t_f} dt'\int_{t_i}^{t'}dt'' \,V_I(t') V_I(t'')+\cdots
.\eeq
The above explicit form of $U_I$ shows that the energy conserved by quantum evolution 
(up to the usual quantum uncertainty $\Delta t \,\Delta E \ge \hbar$) are the eigenvalues of $H$.
Let us consider for example a time-independent interaction $V$
and an initial state and a final state which are energy eigenstates with eigenvalues $E_i$ and $E_f$.
Defining $V^f{}_i=\langle ^f | V |_i\rangle$, at first order one has
\beq\label{eq:V1cte}
 |\bAk{^f}{U}{_i}|^2 \simeq  \bigg |\int_0^t dt' ~e^{i(E_f-E_i)t' } V^f{}_i \bigg|^2 = 
\frac{4| V^f{}_i|^2}{|E_f-E_i|^2}\sin^2\frac{(E_f-E_i)t}{2}
\stackrel{t\to\infty}{\simeq} 2\pi t  | V^f{}_i|^2 \delta(E_f-E_i)
.\eeq
This means that energy conservation reads $E_f=E_i$, up to the usual quantum uncertainty $1/(t_f-t_i)$.
Higher order corrections give the usual sum over intermediate quasi-on-shell states.



\subsection{Example: the indefinite-norm two-state system}\label{example}
Let us consider a two-state system: $\ket{_+}$ with positive unit norm, and $\ket{_-}$ with negative unit norm.
Without loss of generality, 
by redefining the relative phase of the two states and adding a constant overall energy,
one can trivially write the most generic self-adjoint Hamiltonian as
\beq 
 H = \frac{1}{2}\bordermatrix{
  & |_+\rangle & |_- \rangle\cr
\langle_+| &  E_R & -iE_I \cr
\langle_-| & iE_I & E_R }=
\frac{1}{2}\bordermatrix{
  & |^+\rangle & |^- \rangle\cr
\langle_+| &  E_R & iE_I \cr
\langle_-| & iE_I & -E_R }
\eeq
having used $\ket{^\pm}=\pm\ket{_\pm}$.
We see that the $H_{nm}$ components are hermitian, unlike the $H_n{}^m$ components.
The eigenvalues of $H$ are $E_\pm = \pm E$ with $E=\sqrt{E_R^2-E_I^2}/2$. The corresponding eigenstates are
\beq \ket{_{E_+}} =  \sqrt{\frac{\gamma + 1}{2}} \ket{_+} -i \sqrt{\frac{\gamma- 1}{2}} \ket{_-},
\qquad
 \ket{_{E_-}} =i  \sqrt{\frac{\gamma - 1}{2}} \ket{_+} + \sqrt{\frac{\gamma+ 1}{2}} \ket{_-}
 .\eeq
where $\gamma  = 1/\sqrt{1-\sfrac{E_I^2}{E_R^2}} $ is a `boost factor' that substitutes the usual mixing angle.


\begin{itemize}
\item 
If $E_I< E_R$ the eigenvalues of $H$ are real,
the orthogonal eigenvectors satisfy $\bk{_{E_\pm}}{_{E_\pm}} = \pm 1$, 
and tend to get closer to the `light-cone' of zero-norm states as $E_I$ increases.
The components of $U= e^{-i  H t}$ oscillate in time:
\beq 
\label{eq:Unegnorm2}   U=\bordermatrix{
  & |^+\rangle & |^- \rangle\cr
\langle_+| &  \cos(E t)-i\gamma \sin (Et) &  \sqrt{\gamma^2-1}\sin (Et)\cr
\langle_-| & \sqrt{\gamma^2-1}\sin (Et) & \cos(Et)+i \gamma  \sin(Et)
} .
 \eeq
The unusual feature is that $|\bAk{_\pm}{U}{_\pm}|^2$ oscillates between 1 and $\gamma^2\ge 1$.

\item In the critical case, $E_R=E_I$, such that $\gamma=\infty$,
the two eigenstates become degenerate with energy $E=0$.
The two eigenvectors also become degenerate, and parallel to the
zero-norm state $\propto \ket{_+}+i \ket{_-}$.
The evolution operator is
\beq 
   U=\bordermatrix{
  & |^+\rangle & |^- \rangle\cr
\langle_+| & 1-i E_R t/2 &  E_R t /2 \cr
\langle_-| &E_R t/2 & 1+i E_R t /2
} .
 \eeq
This exemplifies a more general result: 
zero-norm eigenstates with complex eigenvalues appear when, increasing the interaction, a level crossing between a positive-norm
and a negative-norm eigenstate takes place;
the Hamiltonian becomes 
degenerate at the critical transition.

\item   If the interaction $E_I$ is strong enough, $E_I >E_R$, one has a pair of complex conjugated eigenvalues, 
with zero-norm eigenvectors that satisfy $\bk{_{E_+}}{_{E_-}}=1$ and describe tachyonic ghosts:  
  their time-evolution 
factor $e^{-iE_\pm t}$ also contains a real exponential, in analogy to tachyonic states present in positive-norm theories.
In the extreme limit $E_I\gg E_R$ the eigenvalues of $H$ are $\pm i E_I/2$, and the
time evolution operator is:
  \beq 
\label{eq:Unegnorm}   
U=\bordermatrix{
  & |^+\rangle & |^- \rangle\cr
\langle_+| &  \cosh(E_I t/2) & \sinh (E_I t/2)\cr
\langle_-| & \sinh (E_I t/2) & \cosh (E_I t/2)
} .
 \eeq

 \end{itemize}
This runaway happens whenever $H$ has a pair of complex eigenvalues $E_+=E_-^*$,
as clear writing time evolution in terms of energy eigenstates,
$|\psi(t)\rangle = \psi^{E_+} e^{-i E_+ t} |_{E_+}\rangle + \psi^{E_-} e^{-i E_- t} |_{E_-}\rangle$.
Both the norm of $\ket{\psi(t)}$
and the real expectation value of $H$ are preserved by time evolution: 
\beq  \frac{ \bAk{ \psi(t) }{ H }{ \psi(t) }}{\bk{ \psi(t) }{ \psi(t) }} = \frac{E_+ \psi^{E_+} \psi^{E_-*} +\hbox{c.c.} }{\psi^{E_+} \psi^{E_-*} + \hbox{c.c.}}.
\eeq




\subsection{The negative-norm harmonic oscillator}\label{osc-}
We here study the concrete system that lies at the basis of perturbative Quantum Field Theory: the harmonic oscillator.
As discussed by Lee and Wick~\cite{LW} it admits two inequivalent quantizations: positive norm, and indefinite norm.

\smallskip

Let us first recall the standard oscillator, described (up to irrelevant constants) by the Hamiltonian
$H = \frac12 (p^2+q^2)$ with $[q,p]=i$.
Defining 
\beq a=\frac{q+ip}{\sqrt{2}} \mio{= \frac{q+\partial_q}{\sqrt{2}}},\qquad a^\dagger =\frac{q-ip}{\sqrt{2}} \mio{= \frac{q-\partial_q}{\sqrt{2}}}\eeq
one has $[a,a^\dagger]=1$ and $H=(a a^\dagger +a^\dagger a) /2$.
\mio{Defining the vacuum as $a\ket{0}=0$, one has the excited states
$\ket{n} = (a^\dagger)^n\ket{0}/\sqrt{n!}$ where the normalization factor is inserted such that
commutation relations imply the norm $\bk{n}{m}=\delta_{nm}$.
The matrix elements of $a$ and $a^\dagger$ then follow from
\beq  a\ket{n}= \sqrt{n} \ket{n-1},\qquad \hbox{so}\qquad
a^\dagger \ket{n} = \sqrt{n+1}\ket{n+1}.\eeq
\mio{Explicitly:
\beq a = \bordermatrix{ & \ket{0} & \ket{1} & \ket{2} & \cdots \cr
\bra{0} & 0&1&0 & \cdots\cr  
\bra{1} &0 &0&\sqrt{2} & \cdots\cr
\bra{2} & 0 & 0 & 0 & \cdots\cr
\vdots & \vdots& \vdots& \vdots&\ddots  },\qquad
a^\dagger= \bordermatrix{ & \ket{0} & \ket{1} & \ket{2} & \cdots \cr
\bra{0} & 0&0&0 & \cdots\cr  
\bra{1} &1 &0&0 & \cdots\cr
\bra{2} & 0 & \sqrt{2} & 0 & \cdots\cr
\vdots & \vdots& \vdots& \vdots&\ddots  }
\eeq}%
The condition $\bAk{q}{a}{0}=0$ is a differential equation that gives the ground-state wave function  $\psi_0(q)\propto e^{-q^2/2}$.
}


\mio{\subsubsection{The negative-norm harmonic oscillator}}

\medskip

Let us next consider a more general system described by the following
Hamiltonian  and commutation relations:
\beq\label{eq:harms}
H = s_H  \frac{a^\dagger a+aa^\dagger}{2} ,\qquad
[a,a^\dagger]=s.\eeq
For $s=s_H=1$ this reduces to the usual oscillator.
We now show that $s=s_H=-1$ defines another consistent positive-energy theory.
The symbol $a^\dagger$ here indicates the adjoint of $a$, which generalizes the
Hermitian conjugate to negative norm.

\smallskip

We again define the vacuum as $a\ket{0}=0$ and the excited states as
$\ket{_n} =a^\dagger\ket{_{n-1}}/\sqrt{n}= (a^\dagger)^n\ket{0}/\sqrt{n!}$.
Its inverse is $a\ket{_n} = s\sqrt{n}\ket{_{n-1}}$.
The state  metric is $\eta_{nm}\equiv \bk{_m}{_n}=s^{n}\delta_{nm}$.
The norm is determined by the dynamics, and odd states have negative norm for $s=-1$.
The inverse metric is $\etat^{nm}=s^{-n}\delta_{nm}$ and the contro-variant states are
$\ket{^n} = s^{-n} \ket{_n}$.
In components one  has:
{\small \beq a = \bordermatrix{ & \ket{0} & \ket{_1} & \ket{_2} & \ket{_3} & \cdots \cr
\bra{0} & 0&s&0 &0& \cdots\cr  
\bra{_1} &0 &0&\sqrt{2}s^2 &0& \cdots\cr
\bra{_2} &0 &0&0&\sqrt{3}s^3 & \cdots\cr
\bra{_3} & 0 & 0 & 0 &0& \cdots\cr
\vdots & \vdots& \vdots& \vdots& \vdots&\ddots  }=
\bordermatrix{ & \ket{0} & \ket{^1} & \ket{^2} & \ket{^3} & \cdots \cr
\bra{0} & 0&1&0 &0& \cdots\cr  
\bra{_1} &0 &0&\sqrt{2} &0& \cdots\cr
\bra{_2} &0 &0&0&\sqrt{3}& \cdots\cr
\bra{_3} & 0 & 0 & 0 &0& \cdots\cr
\vdots & \vdots& \vdots& \vdots& \vdots&\ddots  }
\eeq
{\small \beq a^\dagger = \bordermatrix{ & \ket{0} & \ket{_1} & \ket{_2} & \ket{_3} & \cdots \cr
\bra{0} & 0&0&0 &0& \cdots\cr  
\bra{_1} &s &0&0 &0& \cdots\cr
\bra{_2} &0 &\sqrt{2}s^2&0&0 & \cdots\cr
\bra{_3} & 0 & 0 & \sqrt{3}s^3 &0& \cdots\cr
\vdots & \vdots& \vdots& \vdots& \vdots&\ddots  }=
s\bordermatrix{ & \ket{0} & \ket{^1} & \ket{^2} & \ket{^3} & \cdots \cr
\bra{0} & 0&0&0 &0& \cdots\cr  
\bra{_1} &1 &0&0 &0& \cdots\cr
\bra{_2} &0 &\sqrt{2}&0&0& \cdots\cr
\bra{_3} & 0 & 0 & \sqrt{3} &0& \cdots\cr
\vdots & \vdots& \vdots& \vdots& \vdots&\ddots  }
\eeq}\normalsize
In components the commutation relations read 
\beq
[a,a^\dagger]_{nm} = (a\cdot  \eta \cdot a^\dagger - a^\dagger\cdot \eta \cdot a)_{nm}=  s^{n+1} \delta_{nm} = s\eta_{nm}\qquad \hbox{i.e.}\qquad
[a,a^\dagger] = s \sum_n    \ket{_n}\bra{^n} = s 1\eeq
and the Hamiltonian is:
\beq
H_{nm} = (n+\frac12)  s_H  s^{n+1} \delta_{nm}= E_n \eta_{nm}\qquad \hbox{i.e.}\qquad
H = \sum_{n=0}^\infty  E_n  \ket{_n}\bra{^n} \eeq
where $E_n = (n+\frac12) s s_H$ are the Hamiltonian eigenvalues,  $H\ket{_n}=E_n \ket{_n}$.
We see that positive-energy eigenvalues are obtained for $s=s_H=1$ (the usual case with positive $H$ and positive norm),
but also for $s=s_H=-1$ (negative $H$ and negative norm).

\medskip

Concerning the negative-norm case, $s=-1$,
notice that the harmonic oscillator does not predict tachyonic ghosts with zero norm.
Furthermore the matrix elements $a_{n}{}^m$ are not the hermitian conjugates of $(a^\dagger)_n{}^m$,
such that the operators $q=(a+a^\dagger)/\sqrt{2}$ and $p=i(a^\dagger -a)/\sqrt{2}$
are represented by 
matrices $q_n{}^m$ and $p_n{}^m$ which are not Hermitian. 
This is why various authors who look at these matrices improperly speak of `anti-Hermitian' operators.
Nevertheless, $q$ and $p$ are self-adjoint operators.
We will now find their coordinate representation.


\subsection{The negative-norm coordinate representation}\label{secDP}
Starting from the harmonic oscillator, we now describe a more general representation of a pair of canonical coordinate variables $q,p$.
Parity   flips $q\to -q$ and $p\to - p$.
In the harmonic oscillator case, this means $a\to -a$ and $a^\dagger\to  -a^\dagger$: so eigenstates $\ket{_n}$
with even (odd) $n$ are even (odd) under parity.
In the negative norm quantization,  states with odd $n$ also have negative norm.
Going to the coordinate wave-function representation
(we use the notation $x$ for the coordinate, that later will become field space), this means that  the norm is 
\beq\bk{\psi'}{\psi} =  \int dx\, [ \psi^{\prime*}_{\rm even}(x) \psi_{\rm even}(x) -\psi^{\prime*}_{\rm odd}(x)\psi_{\rm odd}(x)]
= \int dx \, \psi^{\prime*}(x)  \psi(-x)  .  \eeq
The corresponding unit operator is $1 = \int dx \kb{-x}{x}$.
Switching to the formalism appropriate for generic norm, one has  the norm $ \bk{_{x'}}{_x} = \delta(x+x')$.   Thereby the controvariant state is
$\ket{^x}=\ket{_{-x}}$ and it
satisfies the usual $\bk{^{x'}}{_x}=\delta(x-x')$.
As already discussed around eq.\eq{psicocov},
a state can be expanded as $\ket{\psi} = \int dx \, \psi(_x)\ket{^x} =\int dx\,\psi(^x)\ket{_x}$ with $\psi(_x)=\bk{_x}{\psi} $
and $ \psi(^{x})  =\bk{^x}{\psi}=\psi(_{-x})$.

\medskip

What is emerging from the harmonic oscillator computation is a more general structure:
a coordinate space representation of a pair $q,p$ of conjugated canonical variables that differs from the usual positive-norm representation
\beq  { q \ket{x} = x \ket{x},\qquad p \ket{x}  = +i \frac{d}{dx} \ket{x}}
\label{eq:usual}
\eeq
which implies $\bAk{x}{p}{\psi} = (- i d/dx)\psi(x)$ so that it satisfies $\bAk{x}{[q,p]}{\psi} = i \bk{x}{\psi}$.

The negative-norm coordinate representation, originally discussed by Dirac~\cite{Dirac} and Pauli~\cite{Pauli}, is 
\beq  \bbox{ q \ket{_x} = i x \ket{_x},\qquad p \ket{_x}  = + \frac{d}{dx} \ket{_x}.}
\label{eq:DP}
\eeq
Although $q$ looks anti-hermitian, taking into account the extra $i$ as well as the negative norm, these unusual features combine to form a self-adjoint $q$:
 \beq \bAk{_{x'}}{q^\dagger}{_x} = \bAk{_x}{q}{_{x'}}^* =  [ix' \delta(x+x') ]^* = ix \delta(x+x') = \bAk{_{x'}}{q}{_x}. \eeq
 This means that $\bAk{\psi}{q}{\psi}=\int dq~\psi^*(-q) iq\,\psi(q)$ is real.   
A similar result holds for $p$.
When acting on wave-functions one has $\bAk{_x}{q}{\psi}=-ix \psi(_x)$ and $\bAk{_x}{p}{\psi} = (+d/dx) \psi(_x)$,
giving the desired $[q,p]=i$ commutator.
Defining momentum eigenstates as
$p\ket{_p}=ip\ket{_p}$ one finds
$\bk{_q}{_p} = e^{ipq}/\sqrt{2\pi}$,
$\bk{_{p'}}{_p} = \delta(p+p')$.
The operator $q$ acts as
$\bAk{_q}{q}{_p} = (-d/dp) \bk{_q}{_p}$.
One can again define 
$\ket{^p} = \ket{_{-p}}$
such that $1=\int dp \, \kb{^p}{_p}$.

\medskip

The $i$ factor that differentiates the usual representation from the Dirac-Pauli representation has an impact on the time-inversion parity.
As usual, a positive energy spectrum demands that the time inversion symmetry is anti-unitary.
Then, in the Dirac-Pauli quantization  $q$ is naturally $T$-odd and $p$ is naturally $T$-even
(while the opposite holds in the usual  quantization, unless $T$ is defined adding ad-hoc extra signs).
This will play a key role in section~\ref{DP}.

\medskip

We are now ready to come back to the harmonic oscillator.  Inserting into the condition
$\bAk{x}{a}{0}=0$ 
the standard positive-norm representation \mio{($q=x$  and $p = -id/dx$)}%
such that
$a = (q+ip)/\sqrt{2} =(x +s\, d/dx)/\sqrt{2}$
gives a
differential equation which implies the ground-state wave function $\psi_0(x)\propto e^{-sx^2/2}$. 
This is normalizable for $s=1$ (positive norm) and non-normalizable for $s=-1$, where $s$ was defined in eq.\eq{harms}.
This problem was emphaized e.g.\ by Woodard~\cite{Woodard} that thereby dismissed the negative norm quantization as purely formal.
However, the problem arises because the positive-norm representation of $q,p$ was used together with the negative-norm oscillator:
the problem is just a manifestation of the inconsistency of the assumptions.
Consistency demands that the negative norm harmonic oscillator must be accompanied by the negative-norm Dirac-Pauli coordinate space representation of the self-adjoint $q,p$ operators, eq.\eq{DP}.
Then, the condition $\bAk{x}{a}{0}=0$ leads to a normalizable wave function for the ground state $\psi_0 \propto e^{-x^2/2}$, as well as for the excited states.
The Dirac-Pauli choice thereby provides a self-consistent description of the negative-norm oscillator.
Furthermore, as discussed in the next section, in the 4-derivative case the Pauli-Dirac representation is demanded by simple considerations.


\begin{table}
$$\begin{array}{c|cc|cc|c}
\hbox{norm} & \bAk{x}{ q}{\psi} & T\hbox{-parity} &  \bAk{x}{p}{\psi} & T\hbox{-parity} & \hbox{harmonic oscillator with $E>0$}\\ \hline
\hbox{positive}  &x\psi(x) & \hbox{even} & -i\,d\psi/d x & \hbox{odd} &\hbox{$ \psi_0(q) \propto e^{-q^2/2}$ and $H = +\frac12 (q^2+p^2)$}\\
\hbox{indefinite}  &-ix\psi(x) & \hbox{odd} & d\psi/d x & \hbox{even} &\hbox{$ \psi_0(q) \propto e^{-q^2/2}$ and $H =-\frac12 (q^2+p^2)$}
\end{array}$$
\caption{\label{DPtab}\em Coordinate representations of a pair of canonical variables $[q,p]=i$, and the associated ground-state wave functions for the positive-energy
harmonic oscillator.}
\end{table}


\section{4 derivatives want Dirac-Pauli}\label{DP}  
As discussed in the previous section, and as summarized in table~\ref{DPtab}, 
quantum mechanics has two faces: a canonical coordinate can be represented
\begin{itemize}
\item[i)] in the usual way with positive norm; 
\item[ii)] in the Dirac-Pauli way, with negative norm, eq.\eq{DP}.
\end{itemize}
As we now show, theories with 4-derivatives want this latter quantization choice 
(that, in the gravitational case, corresponds to a renormalizable theory with positive energy).

\medskip

A  single 4-derivative real coordinate $q(t)$ contains two degrees of freedom.
The Ostrogradski procedure (section~\ref{OCO}) rewrites the theory as a Hamiltonian system of two canonical coordinates,
$q_1=q$ and $q_2=\lambda \dot q$.
The key new feature that arises in 4-derivative theories is that $\dot q$ becomes an extra canonical coordinate.
In the classical theory $q_2$ is just an auxiliary variable, and $\lambda$ is an irrelevant constant: Ostrogradski used $\lambda=1$.

In the quantum theory, $q_1$ and $q_2$ are operators that allow to define the basis $\ket{q_1, q_2}$.
We now show that the usual quantization must be used for $q_1$ and that the
Dirac-Pauli quantization must be chosen for $q_2$, which is equivalent to 
(and more transparent than) fixing an imaginary $\lambda $ and using the canonical representation.

As usual, the operator $q_1=q$ is invariant under time-reversal $t\to - t$,
and thereby it can follow the usual $T$-even representation.
On the other hand, the operator $\dot q$ is $T$-odd, because of the time derivative: 
the time-inversion operator $T$ transforms it as $T \dot q T^{-1}=-\dot q$.
This is the novel key feature.

Taking into account that $T$ is anti-unitary, one can equivalently define a usual $T$-even coordinate $q_2 = \lambda \dot q_1$
by choosing an imaginary $\lambda$.\footnote{Alternative routes lead to the same conclusion.
For example, one can use the $T$-even
$\ddot q$  (instead of $\dot q$)  as second canonical coordinate. 
In the Ostrogradski formalism $\ddot q = - p_2$ is a momentum.
So again one gets a canonical coordinate with unusual $T$-parity (normally a momentum is $T$-odd).
In general, the invariance of the commutation relation $[q_2,p_2]=i$
under the anti-unitary time-inversion implies that $q_2$ and $p_2$ have opposite $T$-parities.
One can switch $q_2\leftrightarrow p_2$ in order to restore their usual $T$-parities: but their commutator changes sign,
implying again negative norm quantization.  This is indeed what happens in the auxiliary variable formalism,
used in various forms  in the literature as an alternative to the canonical Ostrogradski formalism (see e.g.~\cite{HH,Chen,Grinstein}). This formalism is convenient when dealing with quantum field theory instead of quantum mechanics with a finite number of degrees of freedom.
In order to facilitate the contact, we summarize the auxiliary variable formalism below. Restarting from the Lagrangian in eq.\eq{LagO}, we add zero as a perfect square containing an auxiliary variable $a$:
\beq \label{eq:adda}
 \hbox{\sevenrsfs\relax L}
 = \hbox{\sevenrsfs\relax L}+ \frac12 \bigg[ \ddot q +  (\omega_1^2+\omega_2^2)\frac{q}{2} - \frac{a}{2}\bigg]^2.\eeq
Expanding the square cancels both the second-order and the fourth-order kinetic terms for $q$ leaving
\beq  \hbox{\sevenrsfs\relax L} =-\frac{a \ddot q}{2} + (\omega_1^2-\omega_2^2)^2\frac{q^2}{8}  - (\omega_1^2+\omega_2^2)\frac{aq}{4} + \frac{a^2}{8}-V(q).\eeq
Going to the free theory $V=0$, we can diagonalize the kinetic and mass term  through the field redefinition
\beq \label{eq:auxredef}
\left\{\begin{array}{l}
a = \sqrt{\omega_1^2 - \omega_2^2}(\tilde{q}_2 - \tilde{q}_1)\cr
q = (\tilde{q}_2 + \tilde{q}_1)/{\sqrt{\omega_1^2-\omega_2^2}}
\end{array}\right.\qquad\hbox{i.e.}\qquad
\tilde{q}_{1,2} =\frac{q}{2}  \sqrt{\omega_1^2-\omega_2^2}  \mp \frac{a}{2\sqrt{\omega_1^2-\omega_2^2}} 
\eeq
obtaining two decoupled oscillators
\beq \hbox{\sevenrsfs\relax L} = \frac{\dot{\tilde{q}}_2^2 - \omega_2^2 \tilde{q}_2^2}{2} -  \frac{\dot{\tilde{q}}_1^2 - \omega_1^2 \tilde{q}_1^2}{2}  .  \eeq
From its classical solution,
$a = 2 \ddot q +(\omega_1^2+\omega_2^2)q$, we see that $a$ roughly corresponds to the Ostrogradski $p_2$.
Furthermore, inserting such classical solution in eq.\eq{auxredef} one recovers the formalism used in~\cite{HH} 
\beq \tilde{q}_2 = \frac{\ddot q+\omega_1^2 q}{\sqrt{\omega_1^2-\omega_2^2}},\qquad
 \tilde{q}_1 = - \frac{\ddot q+\omega_2^2q}{\sqrt{\omega_1^2-\omega_2^2}}.\eeq
}
However, it is  simpler to forget  the $\lambda$ factors and just declare that the self-adjoint operator $\dot q$ 
is $T$-odd and thereby it follows the $T$-odd Pauli-Dirac representation.
Then, the Ostrogradski Hamiltonian of eq.\eq{HOstro} is $T$-even.
The states satisfy $T\ket{q,\dot q} = \ket{q,\dot q}$ since $\dot q$ has imaginary eigenvalues and since $T$ is anti-unitary.



\smallskip

The strange extra factor of $i$ has been justified from first principles.
A posteriori, it was not so strange.
After all, it is well known  that the self-adjoint spatial gradient is $i\vec \nabla$ rather than $\vec \nabla$. 
In a relativistic theory, one could have guessed that similarly the
self-adjoint time derivative  is  $i\partial/\partial t$ rather than $\partial/\partial t$.
Loosely speaking, while from a classical perspective $\dot q$ was the natural auxiliary variable,
from  a quantum perspective the natural extra coordinate operator  is $i\dot q$.\footnote{The possibility of
converting a non-normalizable wave-function $\psi_0 \propto e^{z^2/2}$ into a normalizable
one by restricting $z=x+iy$ to the imaginary axis, rather than along the real axis, 
was presented as an ad hoc recipe to get something sensible in earlier works by Bender and Mannheim~\cite{BM}.
At the technical level, their approach differs from ours because they added
an $i$ factor to the variable $q$, rather than to $\dot q$.
Our approach follows from general considerations, and has the advantage that
 $q,p$ and thereby the Hamiltonian are self-adjoint
 (although their matrix representations look anti-hermitian), such that the generalization to an interacting theory will be immediate (section~\ref{interactions}).}

Using the Heisenberg representation, one has $q(t) = U^\dagger(t) q(0) U(t)$ and $\dot q = -i[q,H] = U^\dagger(t) \dot q(0) U(t)$ with unitary $U$, 
so $q(t)$ keeps real eigenvalues and $\dot q(t)$ keeps imaginary eigenvalues at any $t$
(these statements are not contradictory, given that $q(t)$ also depends on $p_1(0)$ and $p_2(0)$).

\subsection*{The frequency eigenstates}

We conclude this section by computing what the Dirac-Pauli representation adopted for $q_2=\dot q$
implies for the frequency eigenstates.
We restart from the Hamiltonian eq.\eq{HOstro} and bring it in diagonal form
\beq H = -\frac12 (\tilde p^{ 2}_1 \tilde \lambda^{ 2} +\omega_1^2 \frac{\tilde q_1^{ 2}}{\tilde\lambda^{ 2}}) + \frac12 (\tilde p^{2}_2 + \omega_2^2 \tilde q^{ 2}_2)\eeq
through the canonical transformation
\beq \label{eq:canqp}
q_1 =\frac{ \tilde{q}_2 -  \tilde{\lambda} \tilde{p}_1/\omega_1}{\sqrt{\omega_1^2 - \omega_2^2}},\qquad
{q_2\over\lambda} = \frac{\tilde{p}_2 - \omega_1\tilde{q}_1/\tilde{\lambda}}{\sqrt{\omega_1^2 - \omega_2^2}},\qquad
p_1 = \omega_1\frac{\omega_1 \tilde{p}_2 -\omega_2^2\tilde{q}_1/\tilde{\lambda}}{\sqrt{\omega_1^2 - \omega_2^2}},\qquad
p_2  \lambda= \frac{\omega_2^2 \tilde{q}_2-\omega_1 \tilde{\lambda} \tilde{p}_1 }{\sqrt{\omega_1^2 - \omega_2^2}}.\eeq
which satisfies $q_1 p_1 - q_2 p_2 = \tilde{p}_2 \tilde{q}_2-\tilde{p}_1\tilde{q}_1$. 
\mio{Its inverse is
\beq  \label{eq:canqp'}
\frac{\tilde{q}_1}{\tilde{\lambda}} =\frac{p_1 - q_2\omega_1^2/\lambda}{\omega_1\sqrt{\omega_1^2-\omega_2^2}},
\qquad
\tilde{q}_2 = \frac{q_1 \omega_1^2 - p_2 \lambda}{\sqrt{\omega_1^2-\omega_2^2}},\qquad
\tilde{\lambda} \tilde{p}_1 = \omega_1\frac{q_1 \omega_2^2 - \lambda p_2}{\sqrt{\omega_1^2-\omega_2^2}},\qquad
\tilde{p}_2 =\frac{p_1 - q_2\omega_2^2/\lambda}{\sqrt{\omega_1^2-\omega_2^2}}
.\eeq}
For the sake of generality, we here allow for generic factors $\lambda$ and $\tilde\lambda$.
The non-vanishing commutators, $[\tilde{q}_1,\tilde{p}_1]=i$ and $[\tilde{q}_2,\tilde{p}_2]=i$,
can be rewritten in terms of $\tilde a_{2} =\sqrt{\omega_2/2} ( \tilde{q}_{2}+ i \tilde{p}_{2}/\omega_2)$
and of $\tilde a_1 =\sqrt{\omega_1/2}(\tilde{q}_1/\tilde\lambda - i \tilde\lambda \tilde p_1/\omega_1)$
reproducing the Hamiltonian of eq.\eq{Htilde} and the commutators
of eq.\eq{a12comm}.
The ground-state wave function is easily computed
imposing  $\bAk{\tilde{q}_1,\tilde{q}_2}{\tilde a_{1,2}}{0}=0$  finding
 \beq\label{eq:psi0'}
 \psi_0(\tilde{q}_1,\tilde{q}_2) \propto\exp\bigg[-\omega_2 \frac{\tilde{q}^2_2}{2}+\omega_1 \frac{\tilde{q}^2_1}{2\tilde\lambda^{2}}\bigg].\eeq
 For $\tilde{\lambda}=1$ it is not normalizable~\cite{Smilga}.
 It is normalizable if instead  $|\Im\tilde{\lambda}|>|\Re \tilde{\lambda}|$.

\medskip

The Dirac-Pauli representation for $q_2,p_2$ corresponds to imaginary
$\lambda$.
Imposing that $q_2, p_1$ are $T$-odd and that $q_1,p_2$ are $T$-even 
(i.e.\ that $q_2$ and $p_2$ have the unusual $T$ parity)
implies that 
$\tilde q_1, \tilde p_2$ are $T$-odd and that $\tilde q_2,\tilde p_1$ are $T$-even
(i.e.\ that the canonical coordinates of the negative norm mode $\tilde{q}_1$ and $\tilde{p}_1$ have the unusual $T$ parity).
This is obtained for imaginary $\tilde\lambda$.


\medskip

As a check,
let us connect the $q_1,q_2$ basis with the $\tilde{q}_1,\tilde{q}_2$ basis for generic $\lambda$ and $\tilde{\lambda}$.
It is convenient to start from the $T$-odd basis $\tilde q_1,\tilde p_2$, in which the ground state wave function is
 \beq\label{eq:psi0q'1p'2}
 \psi_0(\tilde{q}_1,\tilde{p}_2) \propto\exp\bigg[- \frac{\tilde{p}^2_2}{2\omega_2}+\omega_1 \frac{\tilde{q}^2_1}{2\tilde\lambda^{2}}\bigg].\eeq
Next, the transition to the $T$-odd variables $p_1, q_2$ is simply
 \beq
 \bk{p_1,q_2}{\tilde q_1,\tilde p_2}\propto 
 \delta\bigg( \frac{\tilde{q}_1}{\tilde{\lambda}} -\frac{p_1 - q_2\omega_1^2/\lambda}{\omega_1\sqrt{\omega_1^2-\omega_2^2}}\bigg) 
\delta\bigg(\tilde{p}_2 -\frac{p_1 - q_2\omega_2^2/\lambda}{\sqrt{\omega_1^2-\omega_2^2}}\bigg).\eeq
Inserting the change of variables dictated by the $\delta$ functions into $\psi_0(\tilde{q}_1,\tilde{p}_2)$ one obtains
\beq \label{eq:psi0p1q2}
\psi_0(p_1,q_2)\propto \exp\bigg[ -\frac{p_1^2 + 2\omega_1 \omega_2 p_1 q_2/\lambda -\omega_1 \omega_2 (\omega_1^2+\omega_1\omega_2 +\omega_2^2)(q_2/\lambda)^2}{2\omega_1\omega_2(\omega_1+\omega_2)}\bigg]\eeq
where $q_2$ and $p_1$ are both complex and linked by $\Re p_1 = \omega_1^2 \Re (q_2/\lambda)$ and $\Im p_1 = -\omega_2^2 \Im (q_2/\lambda)$.
$\psi_0$ can be trivially analytically continued to real $p_1,q_2$.  For $\lambda=\pm i$ it remains a bounded Gaussian.
Finally, one performs the Fourier transform from $p_1$ to $q_1$, obtaining from $\psi_0(p_1,q_2)$ the ground state wave function
$\psi_0(q_1,q_2)$, which agrees with eq.\eq{psi0div}.
\mio{\footnote{One can also perform a direct computation, although it is more lengthy.
Considering  $\bAk{q_1,q_2}{X}{\tilde{q}_1,\tilde{q}_2} $ with $X=\{q_1,q_2,p_1,p_2\}$
 and equating the expression obtained acting on the left 
with the expression obtained acting on the right using\eq{canqp}, one gets 4 differential equations that fix 
\beq 
\bk{q_1,q_2}{\tilde{q}_1,\tilde{q}_2} \propto \exp\bigg[ -\frac{i}{\tilde{\lambda}} \tilde q_1\omega_1 (\tilde q_2 - q_1 \sqrt{\omega_1^2-\omega_2^2})
+\frac{i}{\lambda} q_2 ( q_1 \omega_1^2- \tilde{q}_2\sqrt{\omega_1^2-\omega_2^2})\bigg].
\eeq
where all products of $q$ are  $T$-odd.
To transform a generic wave function
$\psi(\tilde{q}_1,\tilde{q}_2)$ into $\psi(q_1,q_2)$ one uses 
\beq \bk{q_1,q_2}{\psi} = \int d\tilde{q}_1 d\tilde{q}_2 \bk{q_1,q_2}{\tilde{q}_1,\tilde{q}_2}\bk{\tilde{q}_1,\tilde{q}_2}{\psi}.\eeq
In particular for the ground state one reproduces eq.\eq{psi0div} for all $\tilde{\lambda}$ 
(that disappears).
However, this latter Gaussian integral is not converging for $\omega_1>\omega_2$.}}
The same equality holds for excited states, that can be computed acting with creation operators on the ground state.

In the limit $\omega_1=\omega_2$ one gets the critical situation described in section~\ref{example}.

\section{Path-integral quantization}\label{pathI}
We now present the path-integral quantization of the same 4-derivative theory.

\subsection{Path-integral for generic norm}
Our generic-norm formalism  makes easy to write down the equivalent path-integral formalism, 
an issue already considered in~\cite{Gross}.
Inserting $1 = \int dq \, |^q\rangle\langle _q|$ at intermediate times $t_m = t_i +m\, dt$
one  has
\beq \langle ^{q_f, t_f}| _{q_i, t_i} \rangle =  \prod_m \int dq_m \langle ^{q_{m+1}, t_{m+1} }| _{q_m, t_m}\rangle .\eeq
Each step $\langle ^{q_{m+1}, t_{m+1} }| _{q_m, t_m}\rangle $ can be evaluated as
\beq  
 \langle ^{q_{m+1}}| e^{-i H dt} | _{q_m} \rangle=
 \int dp_m \,\langle ^{q_{m+1}}| ^{p_m}\rangle   \langle _{p_m} |e^{-i H dt} |_{q_m}\rangle
 = \int \frac{dp_m}{2\pi} e^{i [p_m (q_{m+1}-q_m)-H_{\rm cl} dt] }
 \eeq
having defined  
\beq H_{\rm cl} \equiv   \frac{\bAk{_p}{H}{_q} }{\bk{_p}{_q}}\eeq
and used
$\bk{^q}{^p} = e^{ipq}/\sqrt{2\pi}$ and $\bk{_p}{_q}=e^{-ipq}/\sqrt{2\pi}$.
The final result is the path integral
\beq \langle ^{q_f, t_f} | _{q_i, t_i} \rangle =  \int {Dq\,Dp}~e^{i \int dt [p\dot q - H_{\rm cl}]}
\qquad \hbox{where}\qquad
Dq\,Dp =\lim_{dt\to 0} \prod_m \frac{dq_m dp_m}{2\pi}
\label{eq:pathgenericnorm}
\eeq
and with boundary conditions $q(t_i)=q_i$, $q(t_f)=q_f$.

\subsection{Path-integral for 4 derivative quantum theories}
Applying the generic path-integral of eq.\eq{pathgenericnorm} to the
4-derivative oscillator in the canonical Ostrogradski formalism,  one gets the transition amplitude
\beq\label{eq:pathOstro}
\bk{^{q_{1f},q_{2f},t_f}}{_{q_{1i},q_{2i},t_i}}\propto
 \int Dq_1 Dp_1 Dq_2 Dp_2 \, \exp\bigg[{i \int dt \left[p_1\dot q_1 + p_2\dot q_2 - H_{\rm cl}+ J_1 q_1 + J_2 q_2\right]} \bigg]\eeq
where for generality we added currents $J_{1,2}$ 
(such that acting with functional derivatives with respect to them one can form more general 
matrix elements of time-ordered operators; $J_1$ is $T$-even and $J_2$ is $T$-odd).
The Pauli-Dirac representation for $\dot q$ manifests in two ways:

\begin{itemize}
\item[1)] A propagator with an unusual $-$ in its external state.
\end{itemize}
Rewriting the transition amplitude in the usual positive-norm formalism, it
acquires an usual $-$ sign, becoming $\bk{{q_f,-\dot q_f,t_f}}{{q_i,\dot q_i,t_i}}$. 
In the limit $t_f\to t_i$ one has
$\bk{^{q_f,\dot q_f}}{_{q_i,\dot q_i}} = \delta(q_f-q_i) \delta (\dot q_f-\dot q_i)$,
so that the unusual $-$ sign is equivalent to the Dirac-Pauli negative norm.\footnote{
In the limit $dt = t_f- t_i\to 0$ the classical action becomes
\beq S_{\rm cl} \simeq \frac{6}{dt^3}\bigg(q_f - q_i - dt \frac{\dot q_i + \dot q_f}{2}\bigg)^2 + \frac{(\dot q_f - \dot q_i)^2}{2 \, dt}+\cdots\eeq
which is minimal for a motion with constant speed $(\dot q_i + \dot q_f)/2$.
The classical action satisfies $S(q_f,\dot q_f, t_f ;  q_i ,\dot q_i, t_i) = - S( q_i ,\dot q_i, t_i ;q_f,\dot q_f, t_f )$ as well as
$S_{\rm cl}(q_f,\dot q_f,t_f; q_i,\dot q_i, t_i)=-S_{\rm cl}(q_f,-\dot q_f,t_i; q_f,-\dot q_i, t_i)$.} 
Furthermore, the $T$-odd nature of $\dot q$ is hardwired in the path-integral, as a geometrical feature.
For each trajectory $q(t)$ with boundary conditions $q(t_{i,f}) = q_{i,f}$ and $\dot q(t_{i,f}) =\dot q_{i,f}$
the time-inverted trajectory has the same action and
the following boundary conditions:
\beq q_i \to q_f,\qquad q_f\to q_i,\qquad
\dot q_i\to - \dot q_f,\qquad \dot q_f\to - \dot q_i.\eeq
Thereby the propagator given by the path-integral  satisfies the identity
\beq  \label{eq:Kflip}
\bk{q_f, -\dot q_f,t_f}{q_i, \dot q_i,t_i}=\bk{q_i, \dot q_i,t_f}{q_f,- \dot q_f,t_i}\eeq
which is equivalent to the operator identity $\bk{\psi_f}{\psi_i}=\bk{T\psi_i}{T\psi_f}$ given that $T\ket{q,\dot q,t}=\ket{q,\dot q,-t}$.


\begin{itemize}
\item[2)] An unusual classical Hamiltonian.
\end{itemize}
Inserting the Ostrogradski Hamiltonian of eq.\eq{HOstro} 
in the generic path integral of eq.\eq{pathgenericnorm} one gets the following classical Hamiltonian:\mio{\footnote{Using the $\tilde q_{1,2},\tilde p_{1,2}$ basis one instead has, assuming the Dirac-Pauli representation $\tilde\lambda=-i$,
a real $H_{\rm cl} =\sum_{i=1}^2 (\tilde p_i^2 + \omega_i^2 \tilde q_i^2)/2$
such that one obtains convergent path-integrals equal to those of a 2-dimensional 2-derivative harmonic oscillator:
\beq \bk{^{{\tilde q}_{1f},\tilde q_{2f},t_f}}{_{{\tilde q}_{1i},\tilde q_{2i},t_i}}\propto \int D\tilde q_1 D\tilde q_2\,
\exp\bigg[ i \sum_{i=1}^2  \frac{\dot{\tilde{q}}_i^2 - \omega_i^2 \tilde q_i^2}{2}\bigg].\eeq
As already discussed around eq.\eq{psi0p1q2}, its analytic continuation reproduces $\psi_0(q_1,q_2)$.
\xxx{In che senso $\tilde q_1$ \'e T odd?}}}
\beq  H_{\rm cl}=  \frac{ \bAk{_{p_1,p_2}}{H}{_{q_1,q_2}} }{\bk{_{p_1,p_2}}{_{q_1,q_2}}} = i p_1 q_2 + \frac{p_2^2}{2}+\frac{\omega_1^2+\omega_2^2}{2} q_2^2 + \frac{\omega_1^2\omega_2^2}{2} q_1^2+V(q_1).\eeq
This is the same as eq.\eq{HOstro} with $\lambda=-i$.
$H_{\rm cl}$ can be complex because $q_2,p_2$, in the Dirac-Pauli representation, have complex eigenvalues.\footnote{The classical Hamiltonian is real if one instead
uses the equivalent oscillator basis $\tilde{q}_1,\tilde{q}_2,\tilde{p_1},\tilde{p}_2$ of eq.\eq{canqp}.}
\mio{\xxx{In che senso $\lambda q_2$ diventa T even?}
\xxx{what about complex conjugation??:}}
Thanks to the unusual $i$, it is invariant under time-reversal.


\medskip

Let us now try to evaluate the path-integral.
As usual, one can perform the Gaussian $Dp_1 Dp_2$ path integrals.
The $Dp_1$ path-integral formally gives the Dirac delta function
$\delta(q_2-\lambda \dot q_1)$,
allowing to eliminate the $Dq_2$ path-integral, leaving
\beq \bk{^{q_{f},\dot q_{f},t_f}}{_{q_{i},\dot q_{i},t_i}}\propto
\int Dq \,  \exp\bigg[{i \int dt  \left[\Lag(q) + J_1 q +J_2 \lambda \dot q\right]}\bigg],\eeq
where $\Lag$ coincides with the original 4-derivative Lagrangian.
By partial integration, the source term for $\dot q$ can be transformed into a source for $q$ or for $\ddot q$
(like in the auxiliary-field method).
This computation however has three problems:
\begin{enumerate}
\item  the $Dp_1$ path-integral is, in general, divergent.  Thereby the subsequent result is only formal.
\item the $\delta(q_2-\lambda \dot q_1)$  always vanishes if $q_1$ and $q_2$ are real. Thereby the $Dq_2$ path-integral is only formal.
\item Once interactions are turned on, the Lagrangian admits classical run-away solutions, reflected in the path-integral.
\end{enumerate}
Given that the theory is well defined in the operator formalism, somehow this path integral must have a sense.

\subsection{Euclidean Path-integral for 4 derivative quantum theories}

A sensible path-integral is found by restarting from eq.\eq{pathOstro} and continuing it to Euclidean time, $it =  t_E$,
such that $d q/dt = i \, dq/dt_E$ i.e.\ $\dot q = i q'$.
One gets the Euclidean  path integral
\beq \bk{^{q_{1f},q_{2f},t_{Ef}} }{_{q_{1i}, q_{2i},t_{Ei}}}\propto
\int Dq_1Dq_2Dp_1 Dp_2~\exp\bigg[\int dt_E (i p_1 q'_1 + i p_2 q'_2-H_{\rm cl}  + J_1 q_1+J_2 q_2) \bigg].\eeq
Now the $Dp_1$ integral is convergent and gives $\delta(q_2-q'_1)$, such that the $Dq_2$ path integral just fixes $q_2 = q'_1$.
Next, the remaining terms in $H_{\rm cl}$ are a sum of positive squares so all other integrals are convergent.
Performing them one finds the Lagrangian Euclidean path-integral:
\beq\bbox{\bk{^{q_{f},q'_{f},t_{Ef}} }{_{q_{i}, q'_{i},t_{Ei}}} \propto \int Dq \,  \exp\bigg[- \int dt_E  \left[\Lag_E(q) + J_1 q +J_2  q'\right]\bigg]}\eeq
where the classical Euclidean Lagrangian corresponding to eq.\eq{LagO} is
\beq \Lag_E =\frac12 \bigg(\frac{d^2q}{dt_E^2}\bigg)^2
 + \frac{\omega_1^2+\omega_2^2}{2}
 \bigg(\frac{dq}{dt_E}\bigg)^2+
 \frac{ \omega_1^2 \omega_2^2}{2} q^2 + V(q)
.\eeq

\medskip

Let us now check the result.
The classical free solution is
\beq \label{eq:qclE}
q(t_E) = a_1 e^{-\omega_1 t_E}+a_2 e^{-\omega_2 t_E}+ b_1e^{\omega_1 t_E}+b_2 e^{\omega_2 t_E}.\eeq
It already contains run-away
exponentials, characteristic of any Euclidean theory.  Interactions compatible with the positivity of the action
lead to an equally good path-integral.
By imposing the boundary conditions $q= q' =0$ at $t_{Ei}=-\infty$ and evaluating the classical action,
one finds the normalizable ground-state wave function
\beq \label{eq:psi0E}
\bk{q, q', t_E=0}{0,0,t_E=-\infty}\propto \exp\bigg[-\frac{q^2 \omega_1 \omega_2 +    q^{\prime 2}}{2}  (\omega_1 + \omega_2)  +  q  q'\omega_1 \omega_2\bigg].\eeq
This agrees with the ground-state wave-function $\psi_0(q_1,q_2)$ in eq.\eq{psi0div},
that was computed in the Dirac-Pauli formalism in Minkowski space,
after identifying $q=q_1$ and $q'=q_2$.
In other words, $q' = dq/dt_E = -i dq/dt$ coincides with $q_2$,
as computed for  $\lambda=- i$.
The novel feature introduced by 4-derivatives is that $q'$ must {\em not} be continued
into an imaginary $-i \dot q$ (which would give divergent wave functions), because 
it already describes the $T$-odd variable $q_2$, which contains the Dirac-Pauli $i$ factor of eq.\eq{DP}.
\footnote{
Hawking and Hertog~\cite{HH} found
a non-normalizable Minkowskian wave-function because they expressed eq.\eq{psi0E} in terms of $\dot q$,
which is not the canonical coordinate $q_2$ appropriate for 4-derivative theories.
Their  proposal of integrating out $\dot q$ in the Euclidean before performing the analytic continuation to the Minkowskian is not necessary:
the Minkowskian wave functions are normalizable if the appropriate  analytic continuation is performed.
 }
The final result is that the Minkowskian theory is an unusual analytic continuation of the Euclidean theory.

\section{Interactions, Quantum Field Theory, probability}\label{int}
Summarising, we so far considered a  4-derivative harmonic oscillator.
One might think that we achieved nothing~\cite{Smilga}. After all, a classical 4-derivative harmonic oscillator has no run-away problems, see eq.\eq{q(t)},
given that it splits into two decoupled oscillators, one with negative energy and one with positive energy.
The classical trouble starts when they interact.  In this section we will explain that we have achieved instead something useful in an interacting quantum field theory.

\subsection{Adding interactions}\label{interactions}
The quantum formalism was so far developed for the harmonic oscillator
(which corresponds to the modes of a free 4-derivative quantum field theory), finding that
the quantum theory has a positive energy spectrum and no run-away behaviours.
Adding interactions, 
the quantum interacting inherits all these good properties,
as long as interactions are perturbative and as long as the interacting Hamiltonian $H$  remains self-adjoint.

The second issue was the main obstacle to
past attempts of adding ad-hoc unusual $i$ factors in order to make the quantum free theory consistent~\cite{BM}
(normalizable wave functions and unitary evolution with negative norm and positive energy eigenvalues):
adding extra complex factors can  render interactions complex, ruining the theory~\cite{Smilga}.

In our approach the only extra $i$ factor arose from a principled reason:
$\dot q$ is a $T$-odd coordinate that follows the negative-norm Dirac-Pauli representation.
This satisfies all the properties of quantum mechanics, as generalised to negative norms:
$\dot q$ itself is self-adjoint, like $q$ and $\ddot q$.
Thereby any interaction which is a real function of them is self-adjoint.
Our procedure immediately generalizes to the interacting case
(in agravity~\cite{agravity} all interactions are dictated by general covariance).

The perturbativity assumption means that, as long as the 
energy spectrum of the free oscillator gets slightly distorted by interactions,
the energy eigenvalues will remain real and bounded from below
(strongly interacting theories could also be good;
however they seem not needed for the physical application to agravity~\cite{agravity}).

\medskip

One might worry that, even if all energy eigenvalues are positive, the theory possesses negative-norm states with $\bAk{\psi}{H}{\psi}<0$.
Eq.\eq{UI} shows how transition amplitudes can be computed trough perturbation theory:
we see that the energy eigenvalues are the quantity that enters into conservation of energy.
Thereby a theory where all eigenvalues of $H$ (of $H_0$ in the perturbative expansion) are positive is consistent.
As  usual, perturbative computations can be systematised in terms of the propagator.
 By expressing $q=q_1$ in terms of the annihilation and creation operators $a_{i}$, $a_{i}^\dagger$
through eq.s (\ref{a1}) and (\ref{a2}) and 
using the commutation relations
$[\tilde a_i, \tilde a_i^\dagger]=s_i$
we find the propagator
 \begin{eqnsystem}{sys:prop} 
\langle  0 | T q(t)q (t')|0\rangle 
&=& \langle  0 | \theta(t-t') q(t)q(t')+\theta(t'-t)q(t')q(t) |0\rangle  \\
&=& \frac{1}{\omega_1^2-\omega_2^2} \sum_i \frac{s_i}{2\omega_i} [e^{i \omega_i (t-t')}\theta(t'-t)+e^{i\omega_i (t'-t)}\theta(t-t')]   \\
&=&i \frac{1}{\omega_1^2-\omega_2^2}  \int\frac{dE}{2\pi} \sum_i \frac{s_i \, e^{-iE(t-t')}}{E^2-\omega_i^2+i\epsilon}\\
&=&  \int\frac{dE}{2\pi} \frac{-i\,e^{-iE(t-t')}}{(E^2-\omega_1^2+i\epsilon)(E^2-\omega_2^2+i\epsilon)}.
\end{eqnsystem}
where $\epsilon$ is a small positive quantity and we used $s_1=-1$ and $s_2=1$ in the last step.

\medskip

One might worry that, using the Heisenberg picture,  operators satisfy the time evolution equation $\dot A = -i [A,H]$,
which looks dangerously  similar to the classical equation of motion, as given by Poisson parentheses, which has run-away solutions.
However,  the quantum solutions are equal to the classical solutions only in a free theory.
In general operators are not numbers, and the difference (in particular, the Pauli-Dirac representation)
manifests when non-linear interactions are present.
As well known, the Heisenberg equations are in general solved by $A(t) =U^\dagger (t) A(0) U(t)$.
So, all  good properties of negative norm states found in the Schr\"odinger picture remain valid in the
Heisenberg picture, given that they are equivalent.


\subsection{Extension to quantum field theory}
As well known, a single harmonic-oscillator degree of freedom $q(t)$ is the building block for a field such as $\phi(t,x,y,z)$ or $g_{\mu\nu}(t,x,y,z)$.
The expansion of a field in Fourier modes with given momentum works in the 4-derivative case similarly to the 2-derivative case.
As long as, at the end, we are only interested in $S$-matrix elements, all the detailed structure of the quantum mechanical theory,
such as the wave-functions,
gets hidden behind the commutation relations of eq.\eq{a12comm}, which hold separately for each mode.
The usual $i \epsilon$ prescription for the field propagator dictates that amplitudes can be analytically continued from the Euclidean.
Details will be presented elsewhere.

One would like to claim that quantum field theory inherits all good properties of quantum mechanics also when negative norms are present.
 However, while in quantum mechanics interactions can easily satisfy the condition that avoids
`tachionic ghosts' (namely, the interaction strength between two opposite-norm states must be smaller than their
energy difference as discussed in section~\ref{example}),
any interesting quantum field theory leads 
to situations that might violate this condition.
The simplest situation where this occurs is the decay of a
ghost (for example a massive spin 2 graviton at rest),
which can be degenerate with a multi-particle state (for example two photons going in opposite directions with energy equal to half of the ghost mass), 
such that the interaction, no matter how small, can be smaller than the energy difference.
Actually, the ghost is degenerate with an infinite number of similar states, such that an appropriate limit procedure is needed:
in the positive norm case, entropic considerations allow to interpret this situation as particle decay.
A 4-derivative kinetic term $\Pi(p) =-( p^2 - m_1^2)(p^2-m_2^2)$ acquires a positive imaginary part.
We will explore if `ghost decay' can be interpreted like in~\cite{Grinstein}.

\subsection{Ghost does not play dice?} 
So far we carefully avoided talking about probabilities.

The theory is unitary in a negative-norm  space.
Thereby the only remaining difficulty is assigning an interpretation to states that entangle positive norm components with negative norm components.
The Copenhagen interpretation added an extra ingredient external to the deterministic formalism of quantum mechanics:
the Born postulate, according to which:
\begin{quote}\em
``when an observable corresponding to a self-adjoint operator $A$ is measured in a state $\ket{\psi}$,
the result is an eigenvalue $A_n$ of $A$ with probability 
\beq P_n=\frac{\bAk{\psi}{\Pi_n}{\psi}}{\bk{\psi}{\psi}}\qquad\hbox{where}\qquad 
\Pi_n = \frac{\kb{n}{n}}{\bk{n}{n}}\eeq
is the projector over the eigenstate $\ket{n}$ of $A$''.
\end{quote}
For positive norms, these $P_n$ satisfy the probability rules $0\le P_n\le 1$ and $\sum_n P_n=1$;
the average value of $A$ satisfies $\bAk{\psi}{A}{\psi}/\bk{\psi}{\psi} = \sum_n A_n P_n$.

\bigskip

At the moment we do not have a satisfactory generalisation to indefinite norm.
Even worse, the Born postulate is unsatisfactory by itself, given that it describes a
non-local collapse of the wave-function~\cite{EPR}.
In order to make progress, one needs to operate close to the heart of quantum mechanics.
As well known this presents fatal risks: physicists tend to become philosophers.
We conclude by listing some interpretations of quantum mechanics, equivalent to the Copenhagen interpretation,
which could   lead to a satisfactory indefinitive norm quantum mechanics.
\begin{enumerate}

\item Feynman clarified the ontological basis of the Born postulate:  it agrees with experiments, so  `shut up and compute'.
All experiments have so far been performed with  positive norm states.
The negative norm graviton predicted by agravity is beyond the reach of present experiments. On the one hand, this is good because it means that Einstein's general relativity is recovered at large distances; on the other hand, however,
 we do not have experimental guidance.
Lee and Wick proposed that the interpretation issue is bypassed, given that  in quantum field theory
we can only observe asymptotic states, which are made of positive-norm quanta~\cite{LW}. The Lee-Wick idea may be applied to the gravitational theory proposed by Stelle  \cite{Stelle}, as discussed in \cite{Antoniadis, Hasslacher:1980hd}.

\item Any self-adjoint Hamiltonian $H$ gives unitary evolution with respect to many different norms, since each energy eigenstate
evolves picking just a phase.
Defining ghost parity ${\cal G}$ to be the  metrics in the special basis of energy eigenstates
and $\ket{^\psi} = {\cal G}^{-1} \ket{_\psi}$,
a possible generalization of the Born postulate to generic norm is (see also~\cite{BM})
\beq P_n = \bAk{^\psi}{\Pi_n}{_\psi} \qquad\hbox{where}\qquad 
\Pi_n = \kb{^n}{_n}.\eeq
The example of section~\ref{example} gets converted into normal oscillations with mixing angle $\sin^22\theta=E_I^2/E_R^2$.
However, $\bAk{_\psi}{A}{_\psi}$ is real but does not have a probabilistic interpretation, while
$\bAk{^\psi}{A}{_\psi}$ has a probabilistic interpretation but can be complex.

\item Various authors claim that the Born postulate is just
an emergent phenomenon (somehow like friction)
that follows from the fundamental deterministic equations when applied to complex systems 
 in view of spontaneous decoherence~\cite{deco}.

\item Cramer~\cite{Cramer} proposed a ``transactional interpretation'', claiming that 
EPR non-locality results from a cancellation of advanced and retarded waves,
in a time-symmetric set-up (see also~\cite{ABL}) inspired by the analogous formulation of classical electro-dynamics proposed by Dirac and Feynman-Wheeler.
The $\bk{\psi'}{\psi}$ amplitude in the Dirac-Pauli coordinate representation supports the interpretation as
being the overlap of a wave $\psi$ moving forward in time with a wave $\psi'$ moving backwards in time.  

\end{enumerate}
We plan to further investigate such issues.



\section{Conclusions (so far)}\label{end}
We presented the quantization of 4-derivative theories, finding that a unique structure emerges.
We can summarise it as follows.

Quantum mechanics has its usual visible face, where a coordinate operator $q$ is represented as $q\ket{x}=x\ket{x}$.
But quantum mechanics  also has a hidden face, where $q\ket{x}=ix\ket{x}$, as first pointed out by Dirac and Pauli.
Both $q$ and $p$ of a canonical pair $[q,p]=i$ are self-adjoint in both representations.
The main difference is that the usual representation implies positive norms and $q$ is naturally even under time reflection $T$,
while the DP representation leads to states with indefinite norm and to a 
naturally $T$-odd $q$ (in view of the $i$ factor and of the fact that $T$ is anti-unitary)

The Ostrogradski formulation of a 4-derivative degree of freedom $q(t)$ 
(summarised in section~\ref{OCO}) employs two 
canonical coordinates: $q_1=q$ and $q_2 = \dot q$.
For the first time we have observed that $q_1$, which is  $T$-even, naturally follows the usual representation, while $q_2$ which is  $T$-odd,
naturally follows the Dirac-Pauli negative-norm quantization.
This leads to a sensible quantum theory with positive energies and  normalizable wave-functions, as discussed in section~\ref{DP}.

In section~\ref{QM} we presented a new formalism appropriate for generic-norm quantum mechanics,
introducing `covariant' $\ket{_n}$ and `contro-variant' $\ket{^n}$ basis states.
This clarifies why a self-adjoint linear operator can be represented by a matrix that, in some basis, is not hermitian.
A self-adjoint Hamiltonian leads to unitary time-evolution, in the sense that the negative norm is preserved.
Given that $q$, $\dot q$, $\ddot q$, \ldots are self-adjoint, a Hamiltonian which is a generic real function of them
is self-adjoint, leading to sensible interacting quantum theory provided that one avoids tachyons, an observation that was previously overlooked. 
The usual condition that the theory should be free of tachyons is generalised to negative norm quantum mechanics. 

In section~\ref{pathI} we presented the path-integral formulation of negative-norm quantum mechanics.
The classical Hamiltonian becomes complex.
Another new result of this paper is the proof that the normalizable wave functions found in the operator formalism are recovered from the path-integral
after performing naive manipulations over ill-defined objects and/or analytic continuations.
In particular, the version of the path-integral in Euclidean time $t_E = it$ is well defined, and reproduces the usual wave functions
taking into account that $dq/dt_E$ already coincides with the Dirac-Pauli $\dot q$.

The fact that (1) our approach leads to normalizable wave-functions and (2) these wave-functions can  also be deduced from a well-defined Euclidean path-integral clearly show that the right quantization for $\dot{q}$ is the Dirac-Pauli one.

\bigskip

Two issues must be addressed before that these results can be used to obtain a predictive 
renormalizable quantum theory of gravity:
generalisation to quantum field theory, and generalisation of the Born probabilistic interpretation to negative norms.

\footnotesize

\subsubsection*{Acknowledgments}
This work was supported by the ERC grant NEO-NAT,   by the Spanish Ministry of Economy and Competitiveness under grant FPA2012-32828, 
Consolider-CPAN (CSD2007-00042), the grant  SEV-2012-0249 of the ``Centro de Excelencia Severo Ochoa'' Programme and the grant  HEPHACOS-S2009/ESP1473 from the C.A. de Madrid.
We thank all the colleagues who told us that working with ghosts is equivalent to abandoning physics, and 
Enrique \'Alvarez, Jos\'e R. Espinosa, Antonio Gonz\'alez-Arroyo, Martti Raidal, Hardi Veermae, Sergey Sibiryakov, and Guido Altarelli for useful comments and encouragement.
This was the content of my last conversation with Guido --- the present paper was finalized wondering what Guido would have said.

\footnotesize


\footnotesize


\end{document}